# Laser control in a bifurcating region

D. Sugny*, C. Kontz*, M. Ndong[+], Y. Justum[+], G. Dive ** and M. Desouter-Lecomte[‡1]

*Laboratoire de Physique de l'Université de Bourgogne, Unité Mixte de Recherches 5027 CNRS et Université de Bourgogne, BP 47870, 21078 Dijon, France.

[+]Laboratoire de Chimie Physique, Unité Mixte de Recherches 8000, CNRS et Université de Paris-Sud-11, 91405 Orsay Cedex, France.

**Centre d'Ingéniérie des Protéines, Université de Liège, Sart Tilman B6, B-4000 Liège, Belgium.

‡ Laboratoire de Chimie Physique, Unité Mixte de Recherches 8000, CNRS et Université de Paris-Sud-11, 91405 Orsay Cedex, France and Département de Chimie, Université de Liège, Institut de Chimie B6, Sart-Tilman, B-4000, Liège 1, Belgium.

---

[1] Corresponding author : mdesoute@lcp.u-psud.fr

Abstract

We present a complete analysis of the laser control of a model molecular system using both optimal control theory and adiabatic techniques. This molecule has a particular potential energy surface with a bifurcating region connecting three potential wells which allows a variety of processes such as isomerization, tunnelling or implementation of quantum gates on one or two qubits. The parameters of the model have been chosen so as to reproduce the main features of $H_3CO$ which is a molecule-benchmark for such dynamics. We show the feasibility of different processes and we investigate their robustness against variations of laser field. We discuss the conditions under which each method of control gives the best results. We also point out the relation between optimal control theory and local control.

# I. Introduction.

Control of physico-chemical processes by ultra-short laser pulses remains nowadays an attractive and challenging domain. The aim of this kind of control is to design a laser pulse which drives the system from an initial state to a specific target state or even better, to find laser fields able to perform unitary transformations on molecular qubits. By this way, shaped laser pulses have become new reagents for chemical reactions. Some of the most important experimental contributions to this field have been reviewed recently [1]. On the other hand, different control schemes have been proposed, among others we can cite the Brumer-Shapiro coherent control [2 3], the Tannor-Rice-Kosloff local control approach [4 5], the Rabitz optimum control theory (OCT) based on learning algorithms or closed-loop control procedures [6 7 8 9] or the simulated Raman adiabatic passage (STIRAP) scheme [10 11 12].

This article is devoted to a theoretical analysis of different scenarios based on STIRAP (or extension of this process as f-STIRAP [13]) and OCT by working only in the Infra-Red domain, i.e., without transitions *via* excited electronic states. We consider a two-dimensional model of a bifurcating region in the ground potential energy surface of a polyatomic system. Such a region connects three non equivalent wells. A deep reactant well is connected to a symmetric double-well. One passes from the reactant well to the double basin with a large amplitude bending mode of a migrating hydrogen atom around a given bond. The double-well corresponds to an internal rotation of this atom around the axis defined by the particular bond. This three-well bifurcating region is an interesting pinball topography which suggests different processes of control :

- transformation of a delocalized state into a localized state in the double-well potential [14]
- transformation of a localized state of the double-well potential into the other. This has been already proposed in the spirit of Cope rearrangement [15] or enantiomer selection [16]
- isomerization from the reactant well to a given basin of the surface like in hydrogen transfer in organic molecules [17]. In our case, this reaction involves a break of symmetry.

- realization of one or two qubits systems [18 19 20 21 22 23]. The double-well region offers different possibilities for the choice of the quantum numbers which allow to define the qubits (parity or excitation) [24]

We address different control issues: the efficiency of various strategies which depend on the shape of the dipolar surface or, equivalently on the structure of the dipolar matrix and the robustness of the control with respect to the process used and the duration of the pulse. In each case, we also analyse the different pathways which are enforced by the laser field. Finally we briefly discuss the relationship between local control and OCT in the particular case where the objective is to maximize the average value of the projector on a superposed state.

**II. Model**

We consider a model recently proposed which reproduces the main features of a bifurcating region connecting three potential energy wells [25]. Isoenergy contours are presented in Fig. 1. The model is calibrated on an *ab initio* computation at the QCISD level [25] of the isomerization of the methoxy radical into hydroxymethyl which is a molecule benchmark for such energy landscapes [26]. We should emphasize that we do not intend to control dynamics of this particular radical moiety but, we are rather interested in this particular topography for which a convenient analytical expression has been proposed.

Figure 1 near here

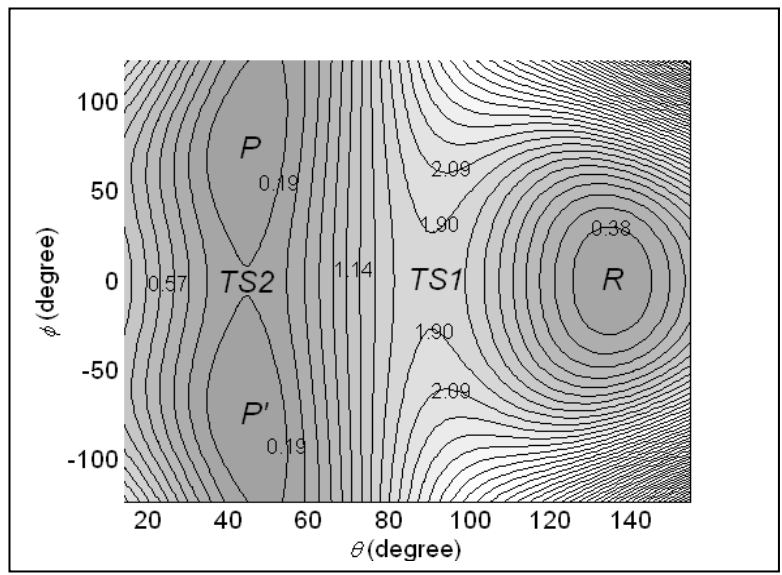

FIG. 1. Isoenergy contours (in eV) in the model potential energy surface of the isomerization $H_3CO \rightarrow H_2COH$ as a function of two active angular coordinates (see Fig.2). The zero of energy is at the bottom of the product well (*P* or *P'*); *R* =0.181 eV, *TS1* = 1.854eV and *TS2* = 0.195eV.

The model describes the rotation of the hydrogen atom around a polar bond connecting two atoms (here CO) in different chemical environments (see Fig.2). The two active coordinates $\theta \in [0, \pi]$ and $\phi \in [-\pi, \pi]$ are the spherical angles of the migrating hydrogen atom with respect to the center of the bond. In $C_S$ geometry ($\phi = 0$), the first active bending coordinate $\theta$ connects the reactant well *R* to a second well trough a first transition state *TS1* (the barrier height from the reactant is 1.673 eV). This second well is a transition state *TS2* according to a second symmetry breaking active coordinate $\phi$. *TS2* is the top of the small barrier (0.195eV) of the double well corresponding to rotational conformers *P* and *P'*. Between *TS1* and *TS2* lies a valley ridge inflexion point (VRI). Mathematical definitions of a VRI point can be found in different works [27 28 29 30]. Roughly speaking, it is a point where a valley corresponding to a particular internal mode becomes an unstable ridge. We point out that field-free dynamics has already been carried out in such bifurcating regions by assuming that initial wave packets can be prepared in the valley uphill from the VRI [31 32]. In the quasi-harmonic regime, one can introduce vibrational quantum numbers for $\theta$ and $\phi$ oscillators. The ground state of the *R* well is denoted by $|0,0\rangle_R$. The delocalized states of the double well are of parity even and odd and are thus noted $|n+, m\rangle$ and $|n-, m\rangle$. The splitting of the first level $|0+,0\rangle$, $|0-,0\rangle$ is $4.3 \ 10^{-5}$ eV. This corresponds to a rather long tunneling time of about 95 ps much longer than the duration of the pulses used in the control. The first localized states coming from the in phase and out of phase superposition are $|nL,0\rangle = (|n+,0\rangle + |n-,0\rangle)/\sqrt{2}$ and $|nR,0\rangle = (|n+,0\rangle - |n-,0\rangle)/\sqrt{2}$. They are associated to *m* = 0 for the $\theta$ vibrator. We recall that the notations R and L do not refer to enantiomers in this example (R is *P* and L is *P'*).

In the dipolar approximation, the reduced 2D Hamiltonian takes the form

$$\hat{H} = \hat{H}_0 - \sum_k \hat{\mu}_k E_k(t) \tag{1}$$

where $\hat{H}_0 = \hat{T} + \hat{V}$ is the field free Hamiltonian and $k$ denotes the polarization direction. The exact constrained 2D kinetic energy operator can be numerically computed by the TNUM algorithm [33] by freezing the inactive coordinates at the *TS1* geometry. We extract an approximate kinetic energy operator by fitting the standard angular momentum expression in spherical coordinates. In Euclidian normalization convention, $\hat{T}$ is then equal to

$$\hat{T}_{Eucl} = -\frac{\hbar^2}{2I_\theta}\left(\frac{\partial^2}{\partial \theta^2} + \cotan\theta \frac{\partial}{\partial \theta}\right) - \frac{\hbar^2}{2I_\phi}\frac{1}{\sin^2\theta}\frac{\partial^2}{\partial \phi^2}$$

where constant inertia moments $I_\theta$ = 6160 a.u. and $I_\phi$ = 4430 a.u. are estimated from the TNUM grids. In the Wilson normalization convention in which the volume element is $d\theta d\phi$, $\hat{T}$ becomes:

$$\hat{T}_{Wil} = -\frac{\hbar^2}{2I_\theta}\frac{\partial^2}{\partial \theta^2} - \frac{\hbar^2}{2I_\phi \sin^2\theta}\frac{\partial^2}{\partial \phi^2} + v(\theta)$$

where $v(\theta)$ is an extra-potential term. This analytical expression is particularly suited to the use of the split operator algorithm [34] which is needed to propagate the wave packets. This point is due to the fact that the coefficient of a given differential operator $\partial/\partial q^k$ does not depend on $q^k$ but only on the other coordinates. We assume that the molecules are aligned in the laboratory frame with the polar bond oriented along the $\vec{e}_z$ axis (see Fig.2). This could be obviously an important constraint [35]. We consider linear polarizations with directions $\vec{e}_x$ in the $C_S$ plane and $\vec{e}_y$ perpendicular to the $C_S$ plane.

Figure 2 near here

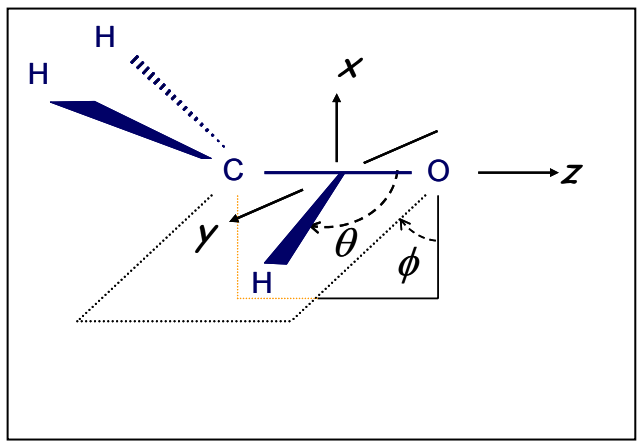

FIG. 2. Active coordinates $\theta$ and $\phi$ for the isomerization $H_3CO \rightarrow H_2COH$ and polarization directions for aligned molecules.

To pursue the construction of the model, we also propose a simple form for dipolar surfaces based on a chemical analysis of the molecule. Indeed, the system can be roughly described as the rotation of a charged particle around a polar bond. The dipolar components $\mu_x(\theta,\phi)$ and $\mu_y(\theta,\phi)$ are larger on the P,P' side ($\theta < \pi/2$) when the particle is close to the most electronegative atom. They decrease quickly for $\theta > \pi/2$, when the particle enters a region near the weakly electronegative atom. The analytical model is given in the appendix. $\mu_x(\theta, \phi)$ and $\mu_y(\theta, \phi)$ are respectively symmetric and antisymmetric with respect to $\phi$. Some cuts are given in Fig.3.

Figure 3 near here

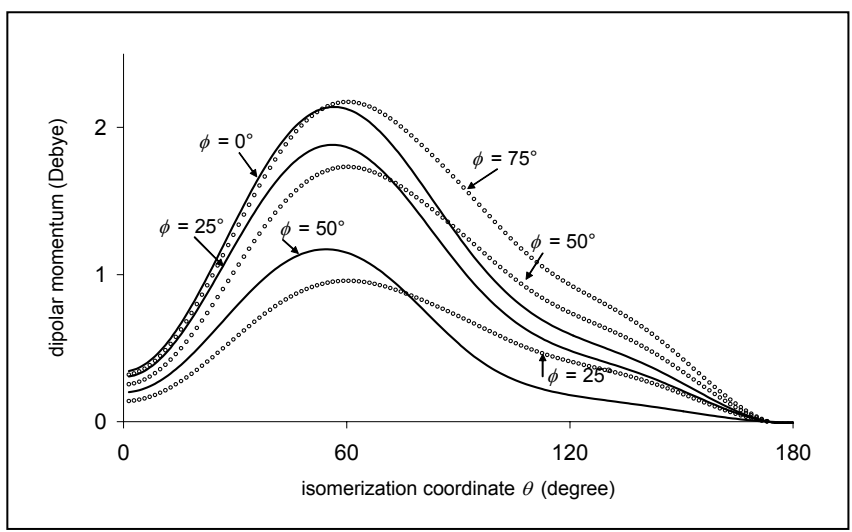

FIG. 3. Cuts in the model dipolar momentum surface (see Appendix) for different values of the torsion angle $\phi$. Full lines : $\mu_x(\theta,\phi)$ ( symmetric), open circles: $\mu_y(\theta,\phi)$ ( antisymmetric).

**III Control methodologies**

**A. Local and optimal methods**

The control algorithms are usually classified as local [4 36 37 38] or global depending on whether the field is determined from the instantaneous dynamical properties by maximizing a performance index or from the variational calculus of a cost functional. The objective functional can be defined in different manners [7 8] which are strongly connected [9]. The procedure to maximize the cost functional under the constraint of satisfying the time dependent Schrödinger equation is described in details in the literature [39]. The Zhu, Botina, Rabitz formulation [7] leads to three coupled equations: the Schrödinger equation for $|\psi(t)\rangle$ with an initial condition $|\psi_i(t=0)\rangle = |\phi_i\rangle$ (forward propagation), the Schrödinger equation for the Lagrange multiplier $|\psi_f(t)\rangle$ with a final target condition $|\psi_f(T)\rangle = |\phi_f\rangle$ (backward propagation) and an equation for the optimum field

$$E_j(t) = -(1/\hbar\alpha_0)\Im m\left[\langle\psi_i(t)|\psi_f(t)\rangle\langle\psi_f(t)|\mu_j|\psi_i(t)\rangle\right] \quad (2)$$

where $\alpha_0$ is a positive penalty factor chosen to weight the significance of the laser fluence. An experimental switching function $s(t) = \sin^2(\pi t/T)$ is usually introduced [39], $\alpha_0$ is then replaced by $\alpha_0 \to \alpha_0/s(t)$. The equations are solved by an iterative formulation [7] adapted to a discrete implementation based on a second order Split Operator scheme [34]. We have used the improvement proposed in Ref. [40]. At each iteration, the field is given by $E_j^{(k)} = E_j^{(k-1)} + \Delta E_j^{(k)}$ where $\Delta E_j^{(k)}$ is calculated by [Eq.(2)].

It is worthy to note that the local approach is strongly related to the Zhu, Botina and Rabitz approach when the performance index involves a projection on a non stationary state. The local control methodology is overviewed in Ref. [38]. The field is chosen in order to maximize the rate of variation of a performance index $y(t) = y(\langle \hat{O}_j(t) \rangle)$ which is a function of expectation values $\langle \hat{O}_j(t) \rangle = \langle \psi_i(t) | \hat{O}_j(t) | \psi_i(t) \rangle$ of Hermitian operators with $j = 1, N$. In the case where the target operator is a projector on a non stationary wave packet at a final time $T$ : $\hat{O}(T) = |\phi_f\rangle\langle\phi_f|$, the rate depends on a single expectation value $dy(t)/dt = d\langle \hat{O}(t) \rangle / dt$. If the time dependence of the operator is fixed by the field free Hamiltonian

$$\hat{O}(t) = e^{-iH^0(t-T)/\hbar} |\phi_f\rangle\langle\phi_f| e^{iH^0(t-T)/\hbar} = |\phi_f(t)\rangle\langle\phi_f(t)|, \tag{3}$$

in other words, when the operator projects down to the wave packet which freely evolves towards the target state at time $T$, then one obtains [38]

$$dy(t)/dt = -2\Im m\left(\langle \hat{O}(t)\hat{\vec{\mu}} \cdot \vec{E}(t) \rangle\right). \tag{4}$$

The local control field giving a monotonous increase of the performance index is obtained by setting, for a polarization direction, $E_j(t) = -\lambda_j \Im m\left(\langle \hat{O}(t)\hat{\mu}_j \rangle\right)$. By inserting expression (3) into this last equation, one gets an expression corresponding to the first step (without zero order field) of the iterative optimum control [Eq.(2)], i.e. when $\psi_f(t)$ evolves with the field free Hamiltonian

$$E_j(t) = -\lambda_j \Im m\left(\langle \psi_i(t) | \phi_f(t) \rangle \langle \phi_f(t) | \hat{\mu}_j | \psi_i(t) \rangle\right) \tag{5}$$

The method focuses on the $\lambda_j$ coefficient. With few trials, it is possible to find values of $\lambda_j$ providing an acceptable field. The latter is then used as an initial-guess field to continue the iterative optimum control procedure. This speeds up the rate of convergence of the algorithm by finally choosing the best $\alpha_0$.

The optimum field able to steer a set of initial states to a set of target states, i.e. to apply a unitary transformation to the $2^N$ states of $N$ qubits

$$\begin{pmatrix} \phi_f^1 \\ \vdots \\ \phi_f^{2^N} \end{pmatrix} = \hat{U}_{gate} \begin{pmatrix} \phi_i^1 \\ \vdots \\ \phi_i^{2^N} \end{pmatrix}$$

can be obtained by the multitarget generalization of OCT [18 21]. We have to propagate simultaneously a set of $2^N$ wave packets forward in time $\psi_i^n(t=0) = \phi_i^n$ with $n = 1,...2^N$ and a set of $2^N$ Lagrange multipliers wave packets backwards $\psi_f^n(t=0) = \phi_f^n$ with $n = 1,...2^N$. The optimum field is given by a sum of contributions from each state

$$E_j(t) = -(1/\hbar\alpha_0)\Im m \sum_{n=1}^{2^N} \left[ \langle \psi_i^n(t)|\psi_f^n(t)\rangle \langle \psi_f^n(t)|\mu_j|\psi_i^n(t)\rangle \right] \qquad (6)$$

A constraint on the phase of the quantum gate could be added [41].

The fidelity of the quantum gate is measured by

$$F = \left| tr(\hat{U}^+_{gate}\hat{U}_{control}) \right|^2 / 2^N \qquad (7)$$

**B. STIRAP and adiabatic processes**

The second strategy for the control is based on adiabatic passage (for a recent overview, see [11 12] and references therein). Such processes are widely used in a variety of fields, extending from nuclear magnetic resonance and quantum information to atomic and molecular excitations. Adiabatic methods are usually achieved by using a series of intense pulses which can be frequency-chirped, the frequencies and the chirping being adapted to the structure of the energy levels. However, the modification of the shape of the pulse envelope and the chirping rate must be sufficiently slow so as to fulfill adiabatic conditions. One of the most well-known adiabatic processes is the STIRAP excitation which involves a counterintuitive sequence of two pulses in a three-level system, in which the field of the Stokes pulse precedes and overlaps the field of the pump pulse. These adiabatic techniques allow a complete population transfer from an initial state to a target state which can be either a stationary state i.e. an eigenstate of the field free Hamiltonian or a coherent superposition of such states. They are also robust in the sense that they are not sensitive to small variations of laser parameters. Due to these remarkable properties, such processes seem to be particularly suitable for the control of chemical reactions. For instance, they have been applied with success for controlling the isomerization of HCN [42 43]. However, the relevance of adiabatic techniques in a complex system can be questioned. We stress that 100% efficiency of the control is generally ensured only for a subset of levels with particular couplings such as the tripod system. If the molecular system is rich in the energy range considered, the effect of coupling to background states can deteriorate noticeably the population transfer in particular if the background states are resonant or almost resonant with laser fields. Another major drawback of these methods is the duration of the pulses which is longer than the time needed by optimal or local control to reach their objective. This point can be problematic if other concurrent chemical processes with time-scale of the same order occur during the control.

To avoid the preceding problems, we combine in this article adiabatic processes which allow determining a simple form for the overall field and optimization of some parameters of the pulse, leading to a shorter (of the order of few picoseconds) and efficient control. The strategy can be summarized as follows. We first select a subset of levels and we determine an adiabatic process in order to achieve the objective of the control. These levels have to be carefully chosen, as otherwise the value of the electric field is too large. More precisely, we recall that the Rabi frequency $\Omega_{12} = |\mu_{12}|E(t)$ between the states 1 and 2 ($\mu_{12}$ being the matrix element of the dipole moment) must be sufficiently large so as to fulfilled adiabatic conditions. For instance, a standard condition is $\Omega_{12}T >> 1$ where $T$ is a characteristic duration of the pulse, which is the full width half maximum for a Gaussian pulse. In addition, in order to avoid other unwanted chemical processes such as ionization, the intensity of the electric field has to be limited to $10^{14}W/cm^2$ which roughly leads to a minimum of the order of 0.1 a.u. for matrix elements of the dipole moment. In a second step, considering all the levels of the system, we decrease the pulse duration to few picoseconds and we optimize both the intensities and the delay between the different pulses to keep efficient control.

We now describe the computational details of the method. We have used Gaussian pulses, the pulses being polarized in the $\vec{e}_x$ or the $\vec{e}_y$ direction. The field $E(t)$ is equal to the sum of terms of the following form :

$$E_0 \exp(-(t-t_k)^2/2\gamma^2)\cos(\omega_k t + \varphi_k) \tag{8}$$

where $\gamma$, $\omega_k$, $E_0$ and $\varphi_k$ are respectively the width, the frequency, the amplitude and the phase of the pulse. To simplify even more the overall field, we assume that the width and the amplitude are the same for all the pulses (except for the quantum gates). The delay is defined by the difference between the times $t_k$.

**IV Wave packet control**

**A. Double well scenarios**

We consider two control schemes in the double well product region (*P* and *P'*, see Fig.1): the localization of the ground delocalized state into one localized state, in the spirit of the previous control on H$_2$POSH [44] and the transformation from a localized state of one well (*P*) to a localized state of the other well (*P'*) [15, 16]. We schematize these processes as follows

$$|0+,0\rangle \rightarrow \frac{1}{\sqrt{2}}(|0+,0\rangle + |0-,0\rangle) = |0L,0\rangle \tag{9}$$

$$|0L,0\rangle \rightarrow \frac{1}{\sqrt{2}}(|0+,0\rangle - |0-,0\rangle) = |0R,0\rangle \tag{10}$$

*(i) Description of the adiabatic processes*

We first analyze in details the processes $|0+,0\rangle \rightarrow |0L,0\rangle$ or $|0+,0\rangle \rightarrow |0R,0\rangle$ which consist in preparing one of the conformer from a delocalized state. The presentation of the results follows the different steps of the strategy. We begin by selecting the first three levels of the system, that is $|0+,0\rangle$, $|0-,0\rangle$ and $|1+,0\rangle$. The method for determining the adiabatic process consists in using the particular symmetry of the dipole moment. For instance, we recall that $\mu_x$ only couples the levels $|0+,0\rangle$ and $|1+,0\rangle$, the transition $|0-,0\rangle$ to $|1+,0\rangle$ being forbidden. We consider a f-STIRAP scheme which, as the STIRAP technique, only uses two pulses, the pump and the Stokes fields. We choose to fix the frequencies $\omega_k$ and the phases $\varphi_k$ of each pulse as follows:

$$\omega_k = E_{1+,0} - \frac{1}{2}(E_{0+,0} + E_{0-,0}) \tag{11}$$
$$\varphi_k = 0$$

where $E_{0+,0}$ is, for instance, the energy of the level $|0+,0\rangle$. In the three-state basis $|0+,0\rangle$, $|0-,0\rangle$ and $|1+,0\rangle$, the total Hamiltonian can be written as

$$\begin{pmatrix} E_{0+,0} & \alpha\Omega_S \cos(\omega t) & \Omega_P \cos(\omega t) \\ \alpha\Omega_S \cos(\omega t) & E_{0-,0} & \Omega_S \cos(\omega t) \\ \Omega_P \cos(\omega t) & \Omega_S \cos(\omega t) & E_{1+,0} \end{pmatrix} \tag{12}$$

where $\Omega_P$ and $\Omega_S$ are respectively the Rabi frequencies of the pump and the Stokes pulses for the transitions $|0+,0\rangle \to |1+,0\rangle$ and $|0-,0\rangle \to |1+,0\rangle$. The $\alpha$ parameter is equal to $\mu^x_{0+0,0-0}/\mu^y_{1+0,0+0}$. The pump pulse is polarized along the $\vec{e}_x$ direction whereas the Stokes pulse is polarized along the $\vec{e}_y$ one. Note that the Rabi frequencies are chosen real without loss of generality. Using the RWA approximation [12], the Hamiltonian $H_I$ reads in the interaction representation

$$H_I = \begin{pmatrix} 0 & 0 & \Omega_P \\ 0 & 0 & \Omega_S \\ \Omega_P & \Omega_S & 0 \end{pmatrix} \tag{13}$$

where the small detunings $E_{1+,0} - E_{0+,0} - \omega$ and $E_{1+,0} - E_{0-,0} - \omega$ are neglected.

The idea is then to use a f-STIRAP technique. f-STIRAP is an extension of STIRAP which allows the creation of coherent superpositions of states [11]. f-STIRAP is now a well-known process which has already been used in a variety of systems to implement qubit gates or to generate superposed states [45 46]. The process can be schematized by the following diagram

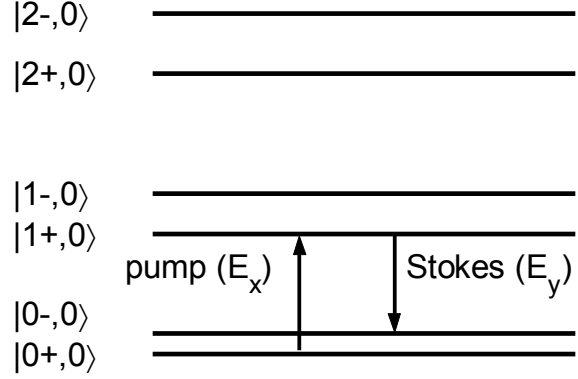

The extension of the STIRAP technique consists in the fact that the amplitudes of the two pulses are required to have a constant ratio at the end of the pulse. More precisely, if the eigenvector $|\psi_0\rangle$ of eigenvalue 0 of $H_I$ writes

$$|\psi_0\rangle = \frac{1}{\sqrt{\Omega_P^2 + \Omega_S^2}} (\Omega_S |0+,0\rangle - \Omega_P |0-,0\rangle) \qquad (14)$$

then the following conditions have to be fulfilled by the two Rabi frequencies:

$$\lim_{t \to -\infty} \frac{\Omega_P}{\Omega_S} = 0 \text{ and } \lim_{t \to +\infty} \frac{\Omega_P}{\Omega_S} = \varepsilon \qquad (15)$$

where $\varepsilon = \pm 1$. One deduces for the two limit cases that:

$$\begin{aligned} |\psi_0(-\infty)\rangle &= |0+,0\rangle \\ |\psi_0(+\infty)\rangle &= \frac{1}{\sqrt{2}}[|0+,0\rangle - \varepsilon |0-,0\rangle] \end{aligned} \qquad (16)$$

It is then clear that for an adiabatic evolution, the localized state can be obtained from f-STIRAP with $\varepsilon = +1$ for $|0R,0\rangle$ and $\varepsilon = -1$ for $|0L,0\rangle$. Another equivalent scheme can be constructed by replacing the intermediate state $|1+,0\rangle$ of the f-STIRAP process with $|2+,0\rangle$:

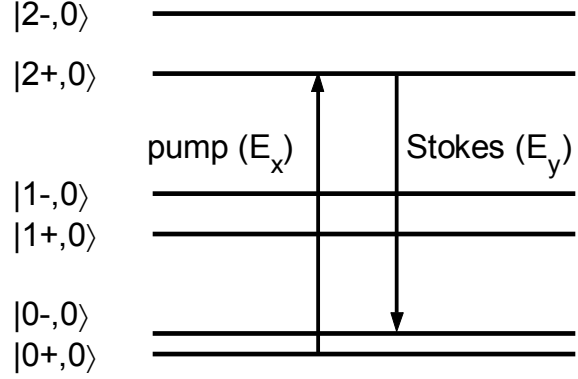

One can also imagine other mechanisms of the same kind using other intermediate states and the particular symmetry of the dipole moment. Finally, the following points can be noticed. A more complex superposed state can be obtained with f-STIRAP if the ratio $\Omega_P/\Omega_S$ is different from 1 or -1 when $t \to +\infty$. Moreover, a process using only one linear polarized laser field but with a frequency $\omega$ and its second harmonic $2\omega$ can also be built to control the tunneling [47].

For the transformation $|0L,0\rangle \to |0R,0\rangle$ from a localized state to the other, we have slightly modified the previous scheme. The limits of [Eq. (15)] become

$$\lim_{t \to -\infty} \frac{\Omega_P}{\Omega_S} = -\varepsilon \text{ and } \lim_{t \to +\infty} \frac{\Omega_P}{\Omega_S} = \varepsilon \qquad (17)$$

leading thus to the following limit states

$$|\psi_0(-\infty)\rangle = \frac{1}{\sqrt{2}}[|0+,0\rangle + \varepsilon|0-,0\rangle]$$
$$|\psi_0(+\infty)\rangle = \frac{1}{\sqrt{2}}[|0+,0\rangle - \varepsilon|0-,0\rangle] \qquad (18)$$

which correspond either to $|0R,0\rangle$ or $|0L,0\rangle$ according to the value of the parameter $\varepsilon$.

Having determined an adiabatic process able to control the specified reaction, we are now in a position to examine the conditions (choice of Rabi frequencies and delay between the pulses) under which the mentioned scheme of control continues to work for a shorter duration of the pulse, i.e. not in the adiabatic limit. The optimized laser field has not been constructed using optimal algorithms in order to preserve as much as possible the robustness of the solution which, as stated above, is one of the most important features of adiabatic processes. For that purpose, we have considered a 2-dimensional grid (Rabi frequencies, delay) and we have calculated for each point of the grid, i.e. for particular values of Rabi frequencies and delay, the corresponding time evolution. The Rabi frequency (the same for each pulse) varies from $10^{-5}$ a.u. to $5.10^{-4}$ a.u. whereas the limits of the delay are $5.10^{4}$ a.u. and $5.10^{5}$ a.u. For our model of $H_3CO$, this amounts to a pulse duration of about 20 ps and a field amplitude of $5.10^{7} Vm^{-1}$ that corresponds to a very weak field.

Figure 4 near here

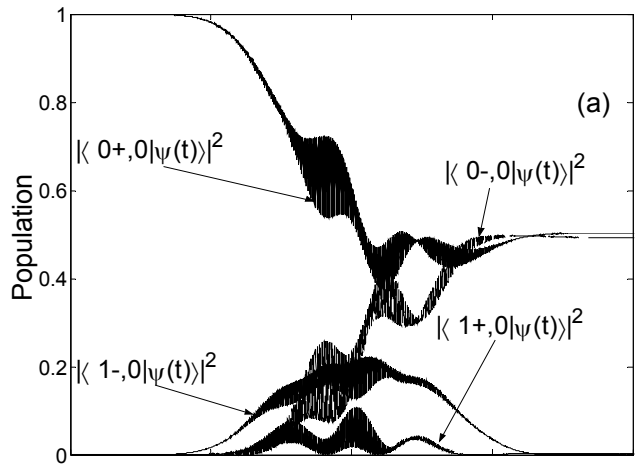

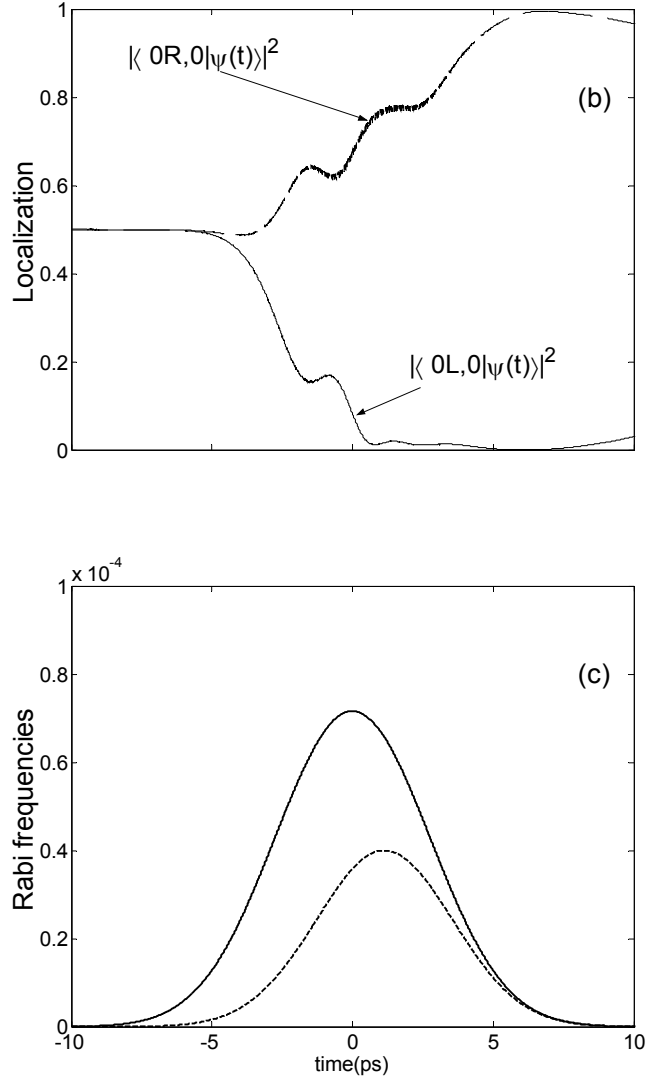

FIG. 4. Dynamics controlled by f-STIRAP strategy for the preparation of the superposed state $|0R,0\rangle$ through the intermediate state $|1+,0\rangle$. Panels (a) and (b) show respectively the evolution of populations in the Hamiltonian eigenbasis and in the superposed states $|0L,0\rangle$ and $|0R,0\rangle$. Populations of other vibrational states remain small during the process. The Rabi frequencies of the different pulses are displayed on panel (c). Rabi frequencies are in atomic units. The solid line corresponds to the Stokes pulse and the dashed line to the pump pulse. The total duration of the pulse is of the order of 20 ps.

Figure 4 illustrates the results of applying the f-STIRAP strategy for a total duration of 20 ps and for the intermediate state $|1+,0\rangle$. In the adiabatic limit, only states $|0+,0\rangle$ and $|0-,0\rangle$ are expected to be populated.

A different behavior is obtained for the process. This is due to coupling to background states and to the fact that adiabatic conditions are not rigorously fulfilled. For instance, the product Rabi frequencies*duration of the pulse is of the order of 10. However, this deviation from the theoretical description does not decrease its efficiency. The same behavior can be obtained for different intermediate states and different total durations.

Figure 5 near here

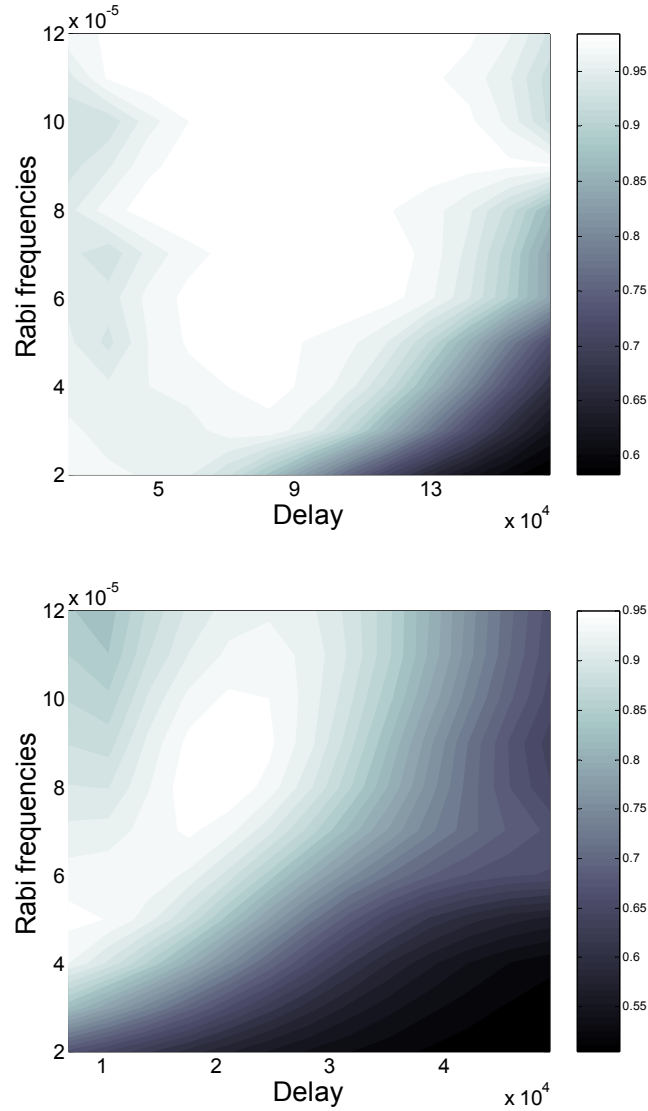

FIG. 5. Robustness of the f-STIRAP process as a function of the Rabi frequencies and the delay between the pulses for a total duration of 20 ps (upper part) and 4.5 ps (lower part) of the overall field. Rabi frequencies and delay are in atomic units. The intermediate state is $|2+,0\rangle$.

The robustness of the strategies has been checked against two parameters: the time delay between successive pulses and the Rabi frequency of each pulse. Figure 5 shows the robustness against these two variables for the f-STIRAP process and for two total durations of about 20 ps and 4.5 ps (see Fig. 4). Remarkable robustness is achieved, especially for the longer pulse duration, advocating for a possible experimental feasibility of the control scheme. Moreover, it can be clearly seen that the strategy is more robust for longer pulses. This point can be explained by the fact that larger the duration of the pulse is, the more the adiabatic conditions are fulfilled and consequently the more the process is robust [48]. The same study with similar results can be done for the transformation from a localized state to the other. However, we notice that the overall dynamics is generally much more oscillatory with more complex structures as compared to the preceding reaction.

*(ii) Optimal control*

Methods based on local or global control allow finding optimum fields with a smaller duration $T$ than fields obtained by the adiabatic approach. Fig. 9 displays the evolution of populations in the eigenbasis of $\hat{H}_0$ [Eq.(1)], in particular for the states $|0+,0\rangle$ and $|0-,0\rangle$ and the average value of the operator $\phi$ showing the localization in the $P'$ well ($\phi = -75°$) at the end of the process, after about 4.5 ps. This time is chosen because it is the shortest time ensuring a good performance index in the STIRAP approach. The objective is reached at 99.99% in 10 iterations with the Rabitz algorithm [Eq.(2)] improved by the correction proposed in Ref.[40] using $\alpha_0 = 1.2$ without any zero order field. Focusing on the first step of the procedure and using local control [Eq.(5)], we obtain a zero order field with $\lambda_x = 8$ and $\lambda_y = 1.2$ leading to a performance index of 91%. The Rabitz algorithm then converges at 99.99% in 3 iterations.

Figure 6 near here

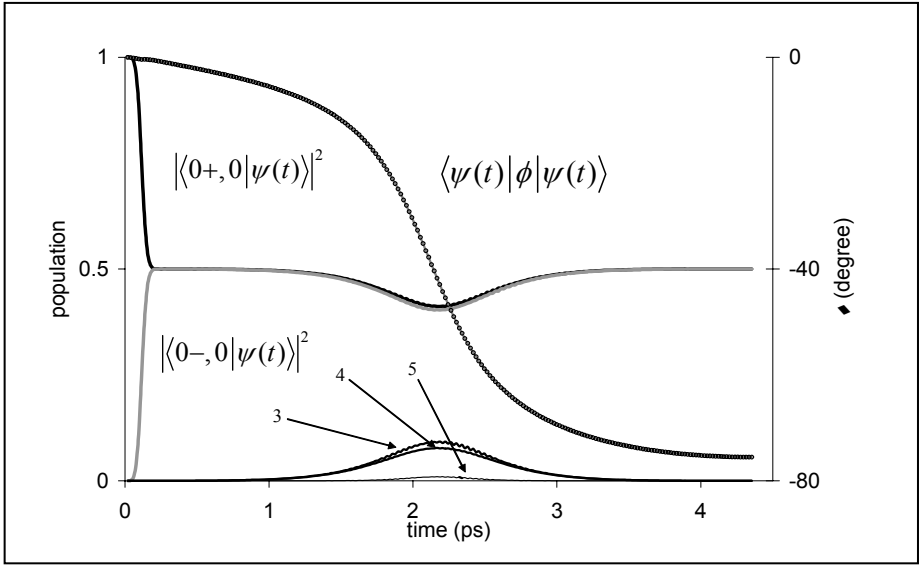

FIG. 6. Dynamics controlled by the optimum field for the preparation of $|0L,0\rangle$. Left axis: evolution of populations in eigenvectors $|0+,0\rangle$ and $|0-,0\rangle$, the excited states 3, 4, 5 nearly correspond to $|1+,0\rangle$, $|1-,0\rangle$ and $|2+,0\rangle$ respectively; right axis: evolution of the average $\phi$ position.

One observes that the populations of the $|0+,0\rangle$ and $|0-,0\rangle$ eigenstates become equal very early but the average $\phi$ position shows that the equality of populations does not involve the correct phase to form the localized superposition $\frac{1}{\sqrt{2}}(|0+,0\rangle+|0-,0\rangle)=|0L,0\rangle$. Transient excitations help at reaching the target superposition. The OCT field is shown in Fig. 7. A part of the structure of the pulse can be understood as follows. The pulse is composed of two sub-pulses, one along the $\vec{e}_x$ direction and the other along the $\vec{e}_y$ direction. We consider this latter part. This sub-pulse can be viewed as a half-cycle pulse (HCP) [49], i.e. only one half of an optical field cycle. HCPs have already been used in different applications; we can cite the control of molecular alignment or orientation [49 50] or the control of tunneling in a double-well system [51]. This pulse, being of short duration with respect to the tunneling time but long in comparison with the period associated to the transitions $|0+,0\rangle \to |1+,0\rangle$ or $|0-,0\rangle \to |1+,0\rangle$, produces a superposition of states $|0+,0\rangle$ and $|0-,0\rangle$. Using the sudden approximation [52], the evolution operator $U_{HCP}$ for the HCP can be written as follows in the basis defined by $|0+,0\rangle$ and $|0-,0\rangle$ :

$$U_{HCP} = \exp(iA_{HCP}\sigma_x) \qquad (19)$$

where $A_{HCP}$ is the area of the pulse times the corresponding matrix element of the dipole moment and $\sigma_x$ the Pauli matrix. Starting from a delocalized state, it can then be shown that a HCP of area $\pi/4$ and a free evolution of a quarter of the tunneling time lead to a completely localized state [51]. A numerical calculation shows that the area of the optimal pulse along the $\vec{e}_y$ direction is very close to $\pi/4$. Notice that the condition $\Omega_{max} \ll \Delta$ where $\Omega_{max}$ is the peak Rabi frequency and $\Delta$ the detuning has to be fulfilled to avoid the appearance of other resonances and the transfer of population to excited states. $\Delta$ can be roughly estimated by $\Delta = E_{1+,0} - E_{0+,0}$ which leads to the following larger possible value of the electric field $E = 2.5 \times 10^{-3}$ a.u. The second part of the pulse along the $\vec{e}_x$ direction is a more complex field which cannot be explained so simply as it does not respect the condition $\Omega_{max} \ll \Delta$. This field is used by OCT to decrease the duration of the control from 20 ps to 4.5 ps.

Figure 7 near here

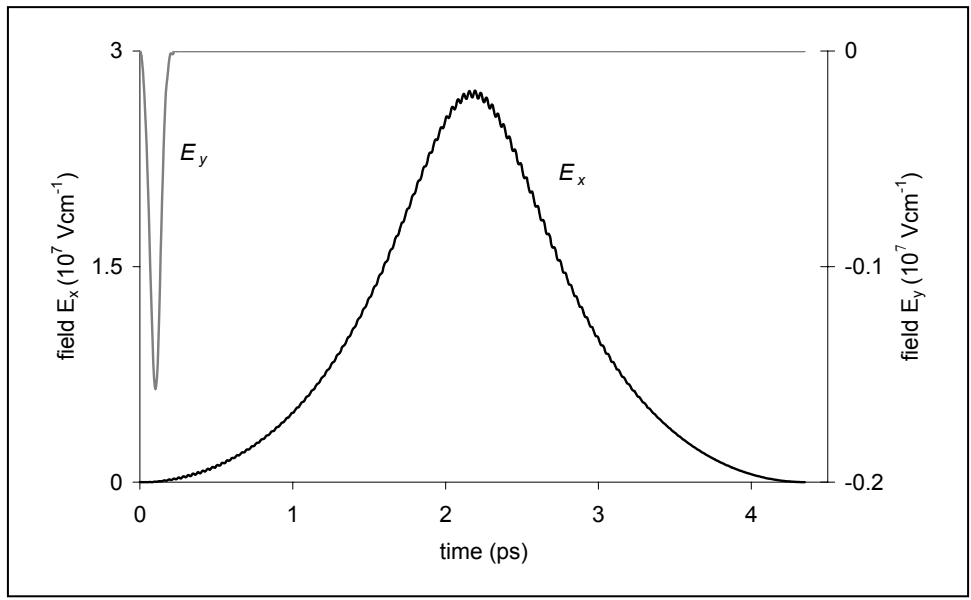

FIG. 7. OCT field obtained for the shortest laser pulse duration studied in the preceding section (f-STIRAP strategy) for the transformation $|0+,0\rangle \rightarrow |0L,0\rangle$ [Eq. (9)].

We have also tested the robustness of this process against the area of the different pulses. As can be seen in [Eq. (19)], the area is the main feature of such short pulses. We have observed that the process is robust (of the order of 10%) with respect to such inaccuracies affecting the area of the pulse. It seems that this feature can be attributed to the simple form of the optimal field.

B. **Bifurcation scenario**

We now investigate the possibility of steering the ground state $|0,0\rangle_R$ of the *R* well towards the *P* $|0R,0\rangle$ or *P'* $|0L,0\rangle$ product basin through the VRI region (see Fig.1). This control is summarized by the following schematic diagram:

$$|0,0\rangle_R \rightarrow |0L,0\rangle \text{ or } |0,0\rangle_R \rightarrow |0R,0\rangle. \tag{20}$$

This scenario involves a break of symmetry after a passage over a high barrier (1.67 eV). From a dynamical point of view, it can be wondered which path the controlled wave packet will follow, i.e. whether the bifurcation occurs early (near the VRI) or not (near *TS2*). It should be noted that this example illustrates the extreme sensibility of the control to parameters of the model. This scenario looks like isomerization processes between two wells which have already discussed in the literature [53 37]. However, the topography between *TS1* and *TS2* with a change of curvature along $\phi$ leads to delocalized eigenvectors strongly coupled by the dipolar momentum. This is unfavourable to the STIRAP scheme which needs intermediate states well decoupled from all the others. In OCT, we do not succeed in finding a satisfactory optimum field for the current model inspired from the QCISD *ab initio* level with a *TS1* barrier of 1.67eV. The algorithm finds a path involving a too high excitation $\langle \Psi(t)|\hat{H}_0|\Psi(t)\rangle$ of several eV up to 7eV which is completely unrealistic. We adopt another potential energy surface inspired from other *ab initio* calculations (MP2) with a smaller energy barrier from the reactant (1.56 eV) at *TS1* ($R$ = 0.451eV, *TS1* = 1.911eV and *TS2* = 0.216eV) and a slightly different profile along $\phi$ for $\phi > 80°$. We use the same dipolar momentum model. In this case, the OCT gives a reasonable field leading to an average unperturbed energy of the order of $E_{TS1}$ as shown in Fig.8. For a pulse duration of 4.5 ps, the target is reached with 95.3% in 240 iterations starting with two zero order fields $E_j^0(t) = E_0 \cos(\omega_j t)$ where $E_0 = 0.02$ a.u. and the $\omega_j$ are the harmonic frequencies of the two $\theta$ and $\phi$ vibrators in the $R$ well. Fig.8 shows the average value of the two active coordinates during the process. The controlled break of symmetry occurs in a sequential manner. The $\theta$ angle first reaches the *TS2* value before the break of symmetry and the cooling occurs in the double well region.

Figure 8 near here

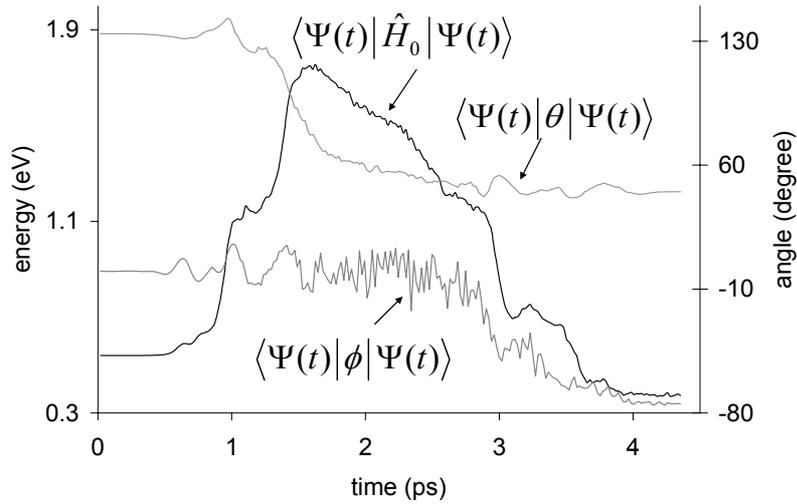

FIG. 8. OCT for the isomerization $R \to P'$. Black line: average unperturbed energy, gray lines: average value of the two active coordinates.

Figures 9 and 10 give the optimal fields and the corresponding Gabor Transforms

$$F(\omega,t) = \left| \int_{-\infty}^{+\infty} H(s-t,\tau)E(s)e^{i\omega s}ds \right|^2 \qquad (21)$$

where $H(s,\tau)$ is the Blackman window [54]

$$H(s,\tau) = 0.08\cos(\frac{4\pi}{\tau}s) + 0.5\cos(\frac{2\pi}{\tau}s) + 0.42 \text{ if } |s| \leq \frac{\tau}{2}$$

$$H(s,\tau) = 0 \text{ elsewhere,}$$

and $\tau$ is the time-resolution. Here we have fixed $\tau = 0.2$ ps.

The Gabor transforms contain the zero order frequencies (1715 cm$^{-1}$ for $E_x$ and 1578 cm$^{-1}$ for $E_y$). A lot of frequencies are used during the time interval [0.8, 1.5] ps. They permit to increase the unperturbed energy above *TS1*. The intermediate states playing a significant role in this heating are the low excitations of the $\theta$ vibrator (nearly $|0,1\rangle_R$, $|0,2\rangle_R$ and $|0,3\rangle_R$) and the first excitation in the $\phi$ vibrator $|1,0\rangle_R$. Between 1.5 ps and 3 ps, a lot of delocalized states are populated with a weight smaller than 5%. At this point, the OCT path involves a large number of intermediate states. Note that this is the extreme opposite of the situation favorable for applying the STIRAP technique. The cooling occurs after 3 ps and mainly involves two states of the double well region (nearly $|0+,1\rangle$ and $|0-,1\rangle$). In this example, cooling is easier than heating probably because the dipolar momentum is very different in the two regions.

Figure 9 near here

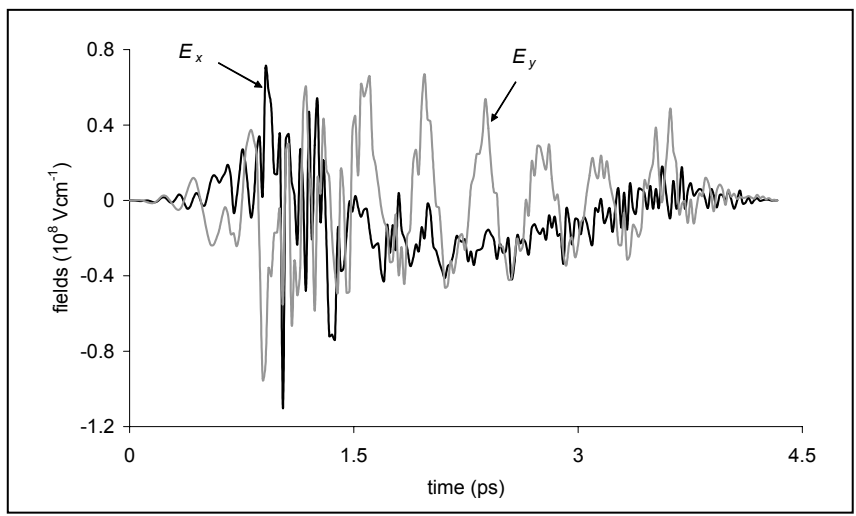

FIG. 9. OCT fields for the izomerization $R \to P'$. Black line: $E_x(t)$, gray lines with open circles: $E_y(t)$

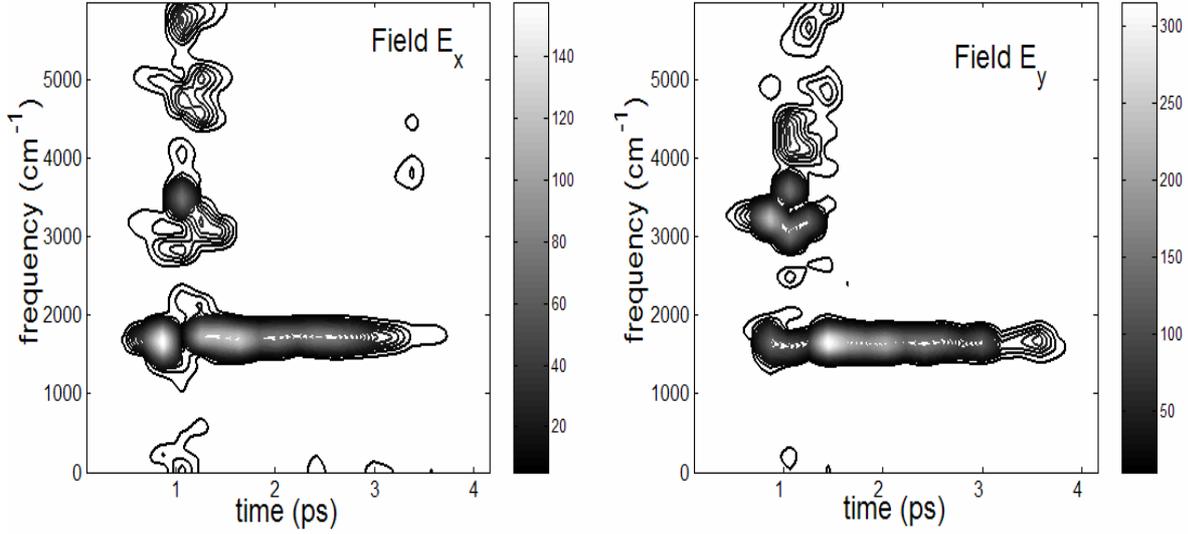

FIG. 10. Gabor transforms [Eq.(21)] of the OCT fields (Fig.9) for the izomerization $R \to P'$.

We have also tried without any success the local approach [37] which consists in heating and cooling the wave packet according to the average $\theta$ position. Besides the difficulty of the expected break of symmetry, the average $\theta$ position does not reach the *TS1* value during the heating but remains in the corresponding well excluding an efficient cooling.

**V. Logical gates**.

We examine the possibility of realizing logical gates on one or two qubits. We recall that a quantum computation can be described as a sequence of logical gates which determine a unitary transformation $U_{gate}$ [55]. Different molecular systems such as vibrationally excited molecules [18 19 20 21 22 23 24] have already been proposed for the implementation of one and two-qubits gates and several control schemes using either pi-pulses [22] or optimal control theory [18 19 20 21 22] have been constructed. In the present case, the low lying states can be thought of as a qubit $|0\rangle = |0+,0\rangle$ and $|1\rangle = |0-,0\rangle$. The previous transformation $|0+,0\rangle \to |0R,0\rangle$ is obviously related to the well known Hadamard transformation

$$\text{HAD}\begin{pmatrix}|0+,0\rangle\\|0-,0\rangle\end{pmatrix}=(1/\sqrt{2})\begin{pmatrix}1&1\\1&-1\end{pmatrix}\begin{pmatrix}|0+,0\rangle\\|0-,0\rangle\end{pmatrix}=\begin{pmatrix}|0L,0\rangle\\|0R,0\rangle\end{pmatrix}.$$

Following this idea, it can be shown that arbitrary unitary operations can be performed on the preceding qubit. For that purpose, we can use a universal set of one-qubit gates composed of the rotation gate and the phase gate which is defined by :

$$\text{PHASE}\begin{pmatrix}|0+,0\rangle\\|0-,0\rangle\end{pmatrix}=\begin{pmatrix}1&0\\0&e^{i\varphi}\end{pmatrix}\begin{pmatrix}|0+,0\rangle\\|0-,0\rangle\end{pmatrix}$$

The basic transformation on a two-qubit system is the controlled-not gate which permutes the state of the second qubit only if the first qubit is in state 1:

$$\text{CNOT}\begin{pmatrix}|00\rangle\\|01\rangle\\|10\rangle\\|11\rangle\end{pmatrix}=\begin{pmatrix}1&0&0&0\\0&1&0&0\\0&0&0&1\\0&0&1&0\end{pmatrix}\begin{pmatrix}|00\rangle\\|01\rangle\\|10\rangle\\|11\rangle\end{pmatrix}$$

There are different ways of defining a two-qubit system in our example. According to the proposal of Sola *et al* [22], we can choose excitation-parity or parity-excitation of the $\phi$ vibrator. This gives, respectively, the following definitions

$$\begin{pmatrix}|00\rangle\\|01\rangle\\|10\rangle\\|11\rangle\end{pmatrix}=\begin{pmatrix}|0+,0\rangle\\|0-,0\rangle\\|1+,0\rangle\\|1-,0\rangle\end{pmatrix}\quad\text{or}\quad\begin{pmatrix}|00\rangle\\|01\rangle\\|10\rangle\\|11\rangle\end{pmatrix}=\begin{pmatrix}|0+,0\rangle\\|1+,0\rangle\\|0-,0\rangle\\|1-,0\rangle\end{pmatrix}$$

We explore also the usual realization of a two-qubit system using states of two $\phi$ and $\theta$ vibrators in the double well

$$\begin{pmatrix}|00\rangle\\|01\rangle\\|10\rangle\\|11\rangle\end{pmatrix}=\begin{pmatrix}|0+,0\rangle\\|0+,1\rangle\\|0-,0\rangle\\|0-,1\rangle\end{pmatrix}.$$

In the next section, we will give examples of control which aim at implementing the Hadamard gate, the phase gate and the C-NOT gate.

**A. One-qubit gate**

*(i) Adiabatic process*

More complex methods than f-STIRAP strategy have to be used for implementing qubit gates. For one-qubit gates, we follow schemes proposed in [56 57]. We only consider the phase gate. The Hadamard gate can be derived by using a similar strategy. The procedure is composed of two STIRAP processes which are aimed at transferring the population between states $|0-,0\rangle$ and $|1+,0\rangle$ via state $|2+,0\rangle$ which is not populated in the adiabatic limit. Only three pulses can be used because the second one serves as a pump field for the first STIRAP excitation and as a Stokes field for the second STIRAP process. The first part of the scheme can be viewed as follows :

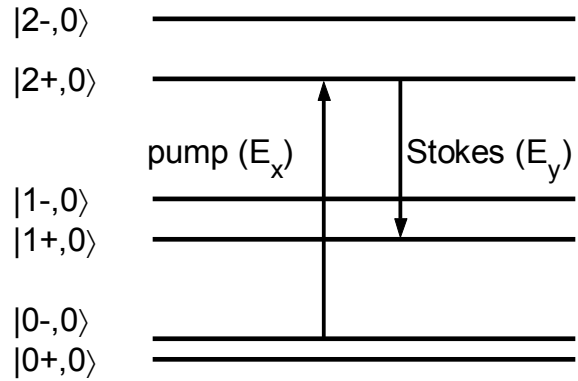

The frequencies are chosen so that the different excitations are resonant, i.e. we have

$$\omega_x = E_{2+,0} - E_{1+,0}$$
$$\omega_y = E_{2+,0} - E_{0-,0}$$

where $\omega_x$ and $\omega_y$ are respectively the frequencies of the fields along the $\vec{e}_x$ and $\vec{e}_y$ directions. The phase of the first pulse is fixed to $\varphi$, the phase of the phase gate, whereas other phases are chosen to be zero. Note that other procedures with different phases can also be considered [56 57]. The eigenvector $|\psi_0\rangle$ of eigenvalue 0 of $H_I$ in the basis $|0-,0\rangle$, $|1+,0\rangle$ and $|2+,0\rangle$ reads as follows during the first STIRAP excitation :

$$|\psi_0\rangle = \frac{1}{\sqrt{\Omega_P^2 + \Omega_S^2}} (\Omega_S |0-,0\rangle - \Omega_P e^{i\varphi} |1+,0\rangle)$$

and the second STIRAP

$$|\psi_0\rangle = \frac{-e^{i\varphi}}{\sqrt{\Omega_P^2 + \Omega_S^2}} (\Omega_S |1+,0\rangle - \Omega_P |0-,0\rangle)$$

In these expressions, the Rabi frequencies are assumed to be real and the dependence on the phase $\varphi$ has been explicitly written in order to clarify the proof.

Figure 11 near here

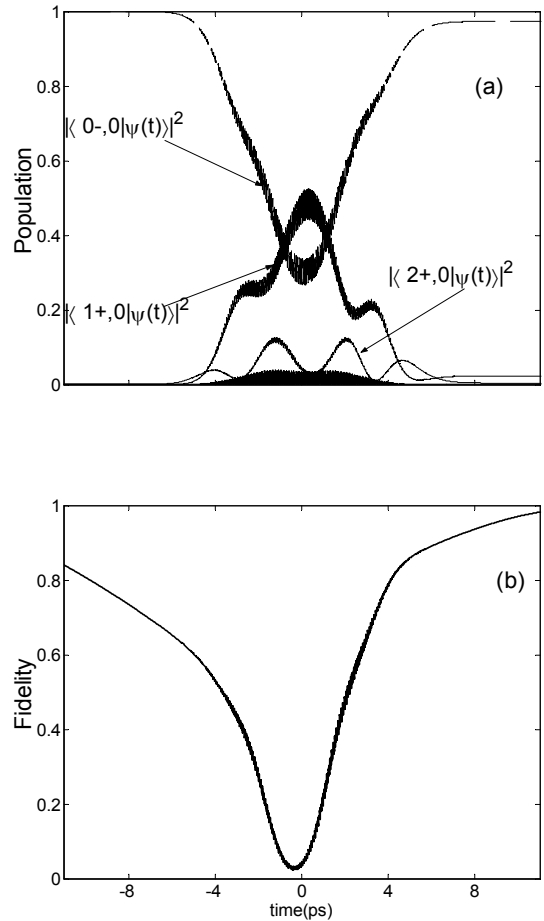

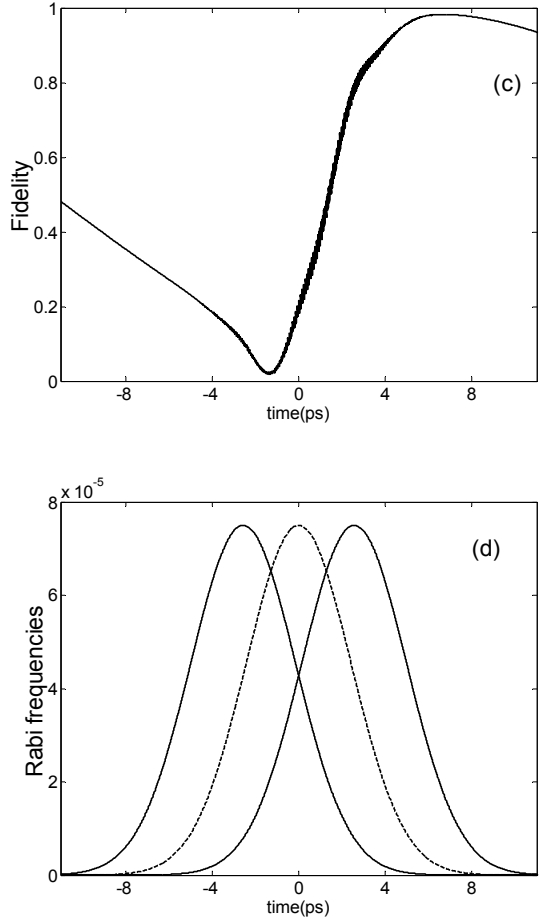

FIG. 11. The phase gate. Panel (a) displays the evolution of populations in the Hamiltonian eigenbasis during the phase gate transformation, the initial state is $|0-,0\rangle$. Panels (b) and (c) represent respectively the evolution of the fidelity for $\varphi = \frac{\pi}{4}$ and $\varphi = \frac{\pi}{2}$. Panel (d) shows the Rabi frequencies of the different pulses. The solid and dashed lines correspond respectively to the field $E_x$ and the field $E_y$ (see text).

It has been more difficult to control and to optimize one qubit gates than processes involved in the double-well scenarios. This point is basically due to the fact that both populations and relative phases have to be controlled in a quantum gate. This difficulty is particularly relevant in this case because the levels of the qubit are not degenerate. The correct unitary transformation is therefore achieved by the adiabatic process only in the interaction representation and not in the bare state basis. The optimization allows to set up the relative phases. Very good results have nevertheless been obtained. Moreover, one of the advantages of adiabatic processes is that the same form of the overall field can be used to realize different quantum gates. This point is illustrated for the phase gate in Fig. 11. Modifying only the phase of the first pulse of the first STIRAP excitation and keeping constant other parameters, two phase gates for $\varphi = \frac{\pi}{4}$ and $\varphi = \frac{\pi}{2}$ have been built by our strategy.

*(ii) Optimal control*

OCT confirms its efficiency in order to find fields of smaller duration. We have derived a field for the Hadamard gate on the $|0+,0\rangle$ and $|0-,0\rangle$ states with $T = 4.5$ ps. The field is completely similar to the one used to realize the scheme (9). One obtains exactly the same behavior as shown in Fig.6. This field is very simple and has been discussed in Sec. IV.A. We have also checked that OCT can be used to implement the phase gate.

**B. Two-qubit gate**

*(i) Adiabatic process*

We only consider the first C-NOT gate involving the states $|0+,0\rangle$, $|0-,0\rangle$, $|1+,0\rangle$ and $|1-,0\rangle$. Similar processes can be constructed for other choices of qubits. The control scheme that can be used for the C-NOT gate is a strategy similar in its spirit to the preceding process. We make a step further by considering now a superposition of states. The scheme can be represented as follows :

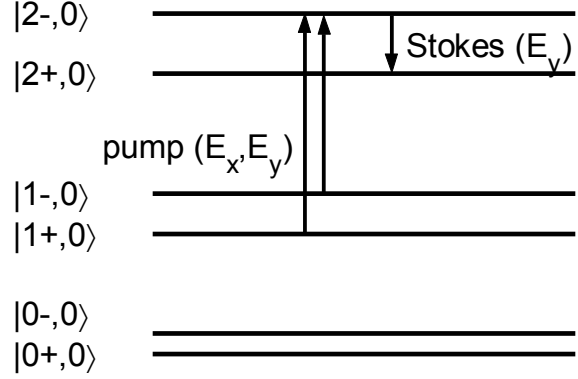

where three different pulses have been considered. The values of the frequencies for the pump pulses and the Stokes field are chosen resonant with the corresponding transition. We consider the subset of levels $|1+,0\rangle$, $|1-,0\rangle$, $|2+,0\rangle$ and $|2-,0\rangle$. In this basis, the total Hamiltonian $H_I$ can be written as

$$\begin{pmatrix} 0 & 0 & 0 & \Omega_{1+} \\ 0 & 0 & 0 & \Omega_{1-} \\ 0 & 0 & 0 & \Omega_S \\ \Omega_{1+} & \Omega_{1-} & \Omega_S & 0 \end{pmatrix}$$

where the Rabi frequencies (with straightforward notations) are assumed to be real. Diagonalizing the matrix C-NOT, we determine the corresponding eigenvectors involving the states $|1+,0\rangle$ and $|1-,0\rangle$. These eigenvectors denoted $|h_+\rangle$ and $|h_-\rangle$ of eigenvalues 1 and -1 can be defined as follows:

$$\begin{cases} |h_+\rangle = \frac{1}{\sqrt{2}}(|1+,0\rangle + |1-,0\rangle) \\ |h_-\rangle = \frac{1}{\sqrt{2}}(|1+,0\rangle - |1-,0\rangle) \end{cases}$$

In the basis $|h_+\rangle$, $|h_-\rangle$, $|2+,0\rangle$ and $|2-,0\rangle$, $H_I$ is given by

$$\begin{pmatrix} 0 & 0 & 0 & \Omega_+ \\ 0 & 0 & 0 & \Omega_- \\ 0 & 0 & 0 & \Omega_S \\ \Omega_+ & \Omega_- & \Omega_S & 0 \end{pmatrix}$$

where a straightforward calculation leads to the following relations :

$$\begin{cases} \Omega_+ = \dfrac{1}{\sqrt{2}}(\Omega_{1+} + \Omega_{1-}) \\ \Omega_- = \dfrac{1}{\sqrt{2}}(\Omega_{1+} - \Omega_{1-}) \end{cases}$$

The idea is then to decouple the eigenvector $|h_+\rangle$ from other states of the basis. For instance, if we choose $\Omega_{1+} = \sqrt{2}\Omega_0$ and $\Omega_{1-} = -\sqrt{2}\Omega_0$, one obtains for $H_I$ :

$$\begin{pmatrix} 0 & 0 & 0 & 0 \\ 0 & 0 & 0 & \Omega_0 \\ 0 & 0 & 0 & \Omega_S \\ 0 & \Omega_0 & \Omega_S & 0 \end{pmatrix}$$

The last step consists in applying the scheme of the phase gate described above for a phase equal to $\pi$. In the adiabatic limit, $|h_+\rangle$ and $|h_-\rangle$ will be respectively transformed into $|h_+\rangle$ and $-|h_-\rangle$ which corresponds to the transformation of the C-NOT gate.

Figure 12 near here

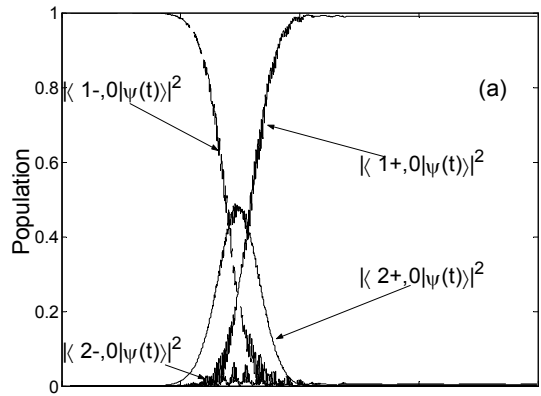

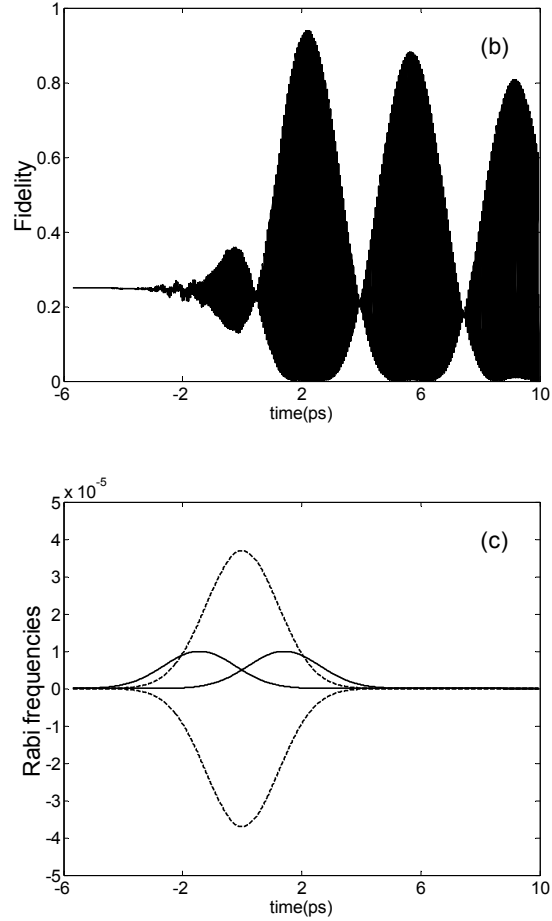

FIG. 12. Same as Fig. 11 but for the C-NOT gate involving the states $|0+,0\rangle$, $|0-,0\rangle$, $|1+,0\rangle$ and $|1-,0\rangle$.

Figure 12 shows the results of this strategy. We have obtained a fidelity close to 0.95. The fact that the levels of the two qubits are not degenerate implies a quick loss of the fidelity of the order of 1 ps which seems problematic in view of experimental applications. We emphasize that this behavior can be observed in most of quantum gates constructed from vibrationally excited states. It is a disadvantage of this kind of systems in comparison of other schemes such as optical cavity [56] where all states are degenerate or almost degenerate.

*(ii) Optimal control*

We present only the gate C-NOT on the two qubits using fundamental and first excited states of two $\phi$ and $\theta$ vibrators in the double well. We impose the pulse duration $T = 4.5$ ps. Fig. 13 displays the time evolution of the population when each initial state $|00\rangle=|0+,0\rangle$, $|01\rangle=|0+,1\rangle$, $|10\rangle=|0-,0\rangle$ and $|11\rangle=|0-,1\rangle$ is driven by the optimum field which has been obtained with 21 iterations. Panels (a) and (b) show the inversion of population of the states of the second qubit. Gray lines display intermediate populations of the different eigenstates. Panels (c) and (d) show the population of the first qubit states. The final value is again equal to one at the end of the process even if intermediate depopulation occurs.

Figure 13 near here

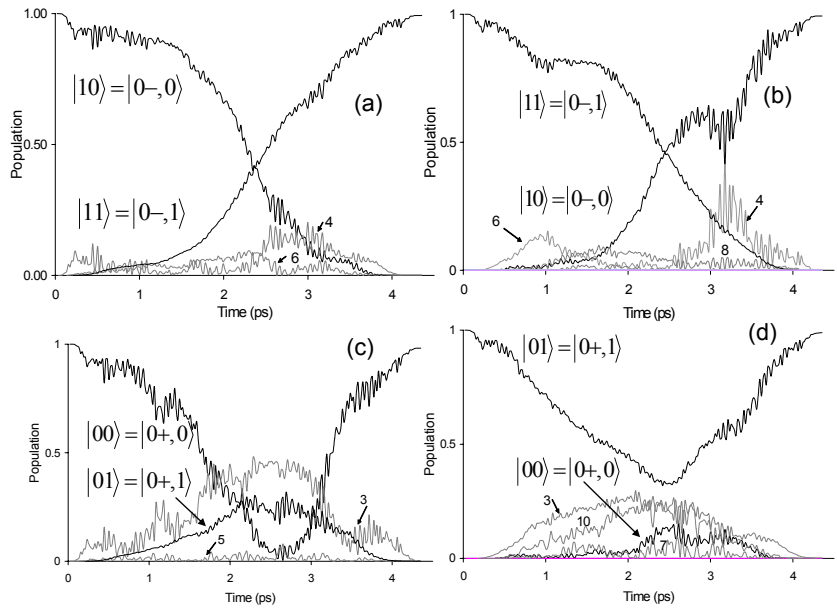

FIG. 13. C-NOT gate on the $|0+,0\rangle,|0-,0\rangle,|0+,1\rangle,|0-,1\rangle$ states. Black lines: population of the states of the two-qubit system, gray lines: intermediate transitions towards other eigenstates which are denoted from 3 to 10 and nearly correspond to excitation of the even and odd states of the $\phi$ vibrator only. The $\theta$ vibrator remains in its ground state with no node along $\theta$.

The optimal field obtained for this C-NOT gate is given in Figure 14. Only the $E_x$ component is used by the OCT. The maximum of the weak $E_y$ component is of the order of $1.5 \ 10^{-3}$ Vcm$^{-1}$. The field is again very simple. The Gabor transform [Eq.(21)] shows that a main frequency 1916 cm$^{-1}$ corresponding to the $|0-,0\rangle \to |0-,1\rangle$ transition acts during the whole process.

Figure 14 near here

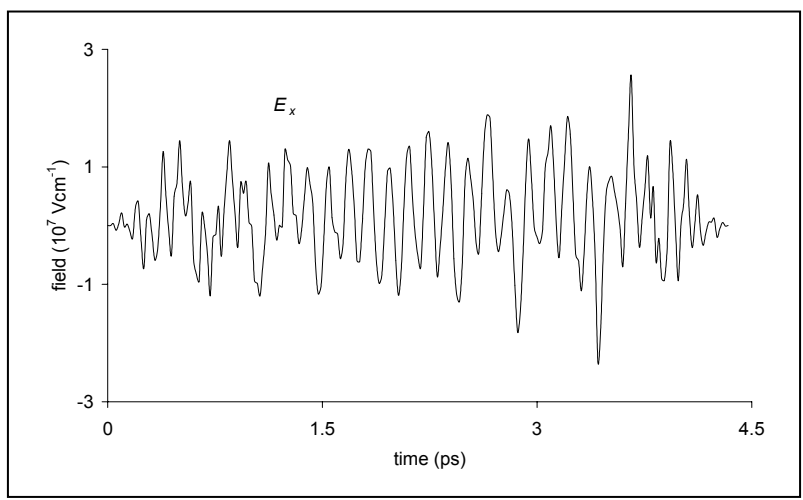

FIG. 14. OCT field for the C-NOT gate on the $|0+,0\rangle, |0-,0\rangle, |0+,1\rangle, |0-,1\rangle$ states.

**VI Concluding remarks**

This article has focused on the application of OCT and adiabatic processes to various situations that can be encountered when a potential energy surface presents a bifurcating region connecting three potential wells (isomerization, tunneling and implementation of one or two qubits quantum gates). In the present case, the symmetric double well region is the most favorable to realize control scenarios due to the shape of the dipolar momentum surface. The results are expected to be transposable to other molecules such as H$_2$POSH presenting the same kind of double well region.

We have also investigated the advantages and limits of the different methods. We recall that the goal of a control is to reach a defined objective, the field solution being subject to some physical constraints: on its duration and its intensity in order to avoid other unwanted chemical processes and on its form and its robustness in view of experimental applications. Some of these constraints are respected by adiabatic processes (simplicity of the form, robustness) and the other by laser pulses determined from OCT (short duration with reasonable intensity). The question which naturally arises is then which strategy used in a given practical situation. Some problems may seem trivial because they consist basically in a jump between two wells. However, as it is the case for our scheme (20), the structure of the eigenvectors and of the dipolar matrix may generate difficulties and hinder the use of STIRAP. The application of adiabatic processes looks particularly problematic if a small number of levels with small coupling to background states cannot be selected. OCT seems to be the more efficient approach even if, as can be shown in Sec.IV.B, there is no guarantee to reach the objective of the control, particularly when the dipole moment is rather flat in a given well. Moreover, if a solution exists there is no more guarantee on the robustness of the optimal field.

For more simple scenarios of control (double-well, qubit gates), several solutions have been obtained, but with different features. In the case of adiabatic processes, one observes that the duration of the overall pulse can be sufficiently reduced (of the order of few picoseconds) by an optimization procedure which also decreases the robustness of the process (Sec.IV.A). For quantum gates where both populations and relative phases have to be controlled, we notice that the reduction of this time and the optimization are more difficult. One of the great advantages of adiabatic processes as compared to OCT is the simplicity of the form of the pulse, the price to pay being generally larger duration and intensity. This is not systematically true as can be shown for the Hadamard gate where a very simple optimal field has been derived (Sec.V.A) by OCT. We have checked the robustness of this latter field and found it to be very good, whereas this is not the case for optimal pulses with more complex structures. Following this example, it seems possible to establish a link between the simplicity of the optimal field and its robustness. We plan to test this conjecture in other molecules in the near future. Finally, we have focused here on dynamics in reduced dimensionality. It is obvious that larger the pulse duration is the more dubious this approximation will be. Our next step will be the consideration of coupling with an environment.

## Appendix

This appendix gives the analytical expression of the dipole moment that has been used in our calculations. We first define the function $\mu_{CS}$ as an approximation of the $\mu_x$ in C$_S$ plane ($\phi = 0$):

$$\mu_{cs}(\theta) = \sum_{k=0}^{4} a_k \cos^k(\theta)$$

where the parameters are given by $a_0 = 0.7$, $a_1 = 1.1$, $a_2 = 0.5$, $a_3 = -1$, $a_4 = -1.11$. The two active coordinates of the dipole moment are then equal to :

$$\mu_x(\theta,\phi) = \mu_{cs}(\theta)\left[f_1(\theta)\cos(\phi) + f_2(\theta)\left(0.25\cos^2(\phi) + 1.75\cos(\phi) - 1\right)\right]$$
$$\mu_y(\theta,\phi) = \mu_{cs}(\theta)\left[f_3(\theta)\sin(\phi)\right]$$

where

$$f_1(\theta) = \frac{1}{2}arctg\left(-3\left(\theta - \frac{\pi}{2}\right)\right) + 0.9$$

$$f_2(\theta) = \frac{1}{2}arctg\left(3\left(\theta - \frac{\pi}{2}\right)\right) + 0.9$$

$$f_3(\theta) = \frac{1}{2}arctg\left(3\left(\theta - \frac{\pi}{2}\right)\right) + 2.4$$

## Acknowledgments

We thank Prof. M. Persico, Prof. H. Jauslin, Dr. S. Guérin, Dr C. Koch, Dr D. Lauvergnat and Dr F.Remacle for helpful discussions. Dr. G. Dive is research associate of the FNRS of Belgium. The computing facilities of IDRIS (Project numbers 061247 and 2006 0811429) as well the financial support of the FNRS in the University of Liège SGI Nic project are gratefully acknowledged.